\documentclass{article}

\usepackage{arxiv}

\usepackage[utf8]{inputenc}
\usepackage[T1]{fontenc}
\usepackage{hyperref}
\usepackage{url}
\usepackage{booktabs}
\usepackage{amsmath,amssymb,amsfonts}
\usepackage{graphicx}
\usepackage{textcomp}
\usepackage{array}
\usepackage[caption=false,font=normalsize,labelfont=sf,textfont=sf]{subfig}
\usepackage{float}
\usepackage{pifont}
\usepackage[]{algorithm2e}
\usepackage{algorithmic}
\usepackage{doi}

\title{An Enhanced Source-Free Unsupervised Domain Adaptation Framework
for Cross-Dataset EEG Emotion Recognition via Predictive Coding
and Test-Time Training}

\renewcommand{\shorttitle}{Enhanced SF-UDA for EEG Emotion Recognition}

\author{
  Md Niaz Imtiaz \\
  Department of Electrical, Computer and Biomedical Engineering \\
  Toronto Metropolitan University \\
  Toronto, ON M5B 2K3, Canada \\
  \texttt{niaz.imtiaz@torontomu.ca} \\
  \And
  Naimul Khan \\
  Department of Electrical, Computer and Biomedical Engineering \\
  Toronto Metropolitan University \\
  Toronto, ON M5B 2K3, Canada \\
  \texttt{n77khan@torontomu.ca} \\
}

\hypersetup{
  pdftitle={An Enhanced Source-Free Unsupervised Domain Adaptation Framework for Cross-Dataset EEG Emotion Recognition},
  pdfsubject={cs.LG, eess.SP},
  pdfauthor={Md Niaz Imtiaz, Naimul Khan},
  pdfkeywords={EEG emotion recognition, source-free domain adaptation,
               self-supervised learning, predictive coding, consistency learning},
}

\begin{document}
\maketitle

% -------------------------------------------------------
\begin{abstract}
EEG-based emotion recognition is widely used in affective computing
but suffers from poor generalization due to domain shifts caused by
inter-subject variability, dataset differences, and recording
conditions, especially in cross-dataset settings. Conventional
unsupervised domain adaptation methods require source data, which is
often unavailable due to privacy constraints. Although source-free UDA
addresses this limitation, existing methods still struggle with large
domain gaps, noisy pseudo-labels, and unstable adaptation. To address
these challenges, we propose an enhanced source-free unsupervised
domain adaptation (SF-UDA) framework for cross-dataset EEG emotion
recognition. The framework introduces a non-contrastive predictive
coding--based self-supervised pretraining strategy to learn robust and
transferable EEG representations by modeling temporal dependencies in
a reconstruction-based manner. During adaptation, we estimate
target-domain structure through class-wise clustering and prediction
disagreement, and optimize the model using a dual-stage adaptation
strategy consisting of Multi-Loss Adaptive Regularization and
Localized Consistency Learning. This dual-stage design improves
stability and neighborhood consistency under noisy pseudo-labels.
Furthermore, we propose a lightweight test-time training mechanism
that enables selective online updates for uncertain samples using
predictive reconstruction loss and entropy minimization, enhancing
robustness during inference while maintaining computational
efficiency. Experiments on the DEAP, SEED, and DREAMER datasets show
consistent improvements over state-of-the-art SF-UDA methods for both
binary and multi-class classification, achieving 69.56\% and 63.03\%
accuracy on SEED and DREAMER when trained on DEAP, and 61.38\% and
68.90\% when trained on SEED. These results demonstrate the
effectiveness of combining predictive self-supervised learning,
source-free domain adaptation, and test-time refinement for robust
EEG emotion recognition under severe domain shifts.
\end{abstract}

\keywords{EEG emotion recognition \and source-free domain adaptation
  \and self-supervised learning \and predictive coding
  \and consistency learning}

% -------------------------------------------------------
\section{Introduction}
\label{introduction}

Emotion is a fundamental aspect of human cognition and behavior,
influencing how people think, communicate, and interact. Consequently,
emotion recognition has become an important research area in
psychology, healthcare, and human--computer interaction. In
healthcare, it can assist in diagnosing and monitoring neurological
and mental health conditions such as autism spectrum disorder (ASD),
depression, and stress-related
disorders~\cite{guo2024development,rashidan2021technology}. It also
plays a key role in affective brain--computer interface (aBCI) systems
that adapt to users' emotional states.

Among various modalities, electroencephalogram (EEG) signals have
attracted significant attention because they directly reflect brain
activity and are less affected by intentional expression or
environmental factors than facial expressions or
speech~\cite{wang2024research}. Their high temporal resolution and the
increasing availability of wearable EEG devices further support their
use in affective computing applications.

Recent advances in deep learning have substantially improved EEG
emotion recognition through automatic feature learning from raw or
transformed EEG signals~\cite{wang2024eeg,kouka2023eeg,fan2024icaps}.
Although convolutional, recurrent, and transformer-based models
achieve strong performance within a dataset, their generalization
ability often deteriorates across datasets due to domain shifts caused
by differences in subjects, recording devices, electrode
configurations, and experimental protocols.

To address domain discrepancies, unsupervised domain adaptation (UDA)
has been widely studied in EEG emotion
recognition~\cite{jimenez2023learning,li2025adversarial,tang2025uda,he2022adversarial}.
However, most existing methods focus on cross-subject or cross-session
adaptation within the same dataset and require access to source-domain
data during adaptation. This assumption is often impractical in
real-world healthcare and brain--computer interface applications,
where sharing raw EEG recordings may be restricted due to privacy,
storage, ethical, and legal considerations.

Source-free unsupervised domain adaptation (SF-UDA) has recently
emerged as a promising alternative. SF-UDA adapts a model to an
unlabeled target domain using only a pre-trained source model,
eliminating the need for source-domain data~\cite{liang2020we}. While
this setting is better suited to privacy-sensitive applications, it
introduces challenges such as source-model bias, the lack of explicit
source-target alignment, and error accumulation from unreliable
pseudo-labels.

Although SF-UDA has achieved encouraging results in computer
vision~\cite{du2024generation,ahmed2023ssda,liu2025srpl,jiang2025hg,huang2022relative,yang2022source},
its application to EEG analysis remains limited. Existing EEG-related
SF-UDA studies mainly focus on seizure and epilepsy detection and
sleep stage
classification~\cite{zhao2023source,ragab2023source,wang2025active},
while research on EEG-based emotion recognition is still limited.
Moreover, many existing methods rely on single-stage adaptation
strategies that struggle to handle the substantial domain
discrepancies encountered in cross-dataset EEG emotion recognition.

To address these challenges, we propose an enhanced source-free
unsupervised domain adaptation framework for cross-dataset EEG-based
emotion recognition, improving upon our previous SF-UDA
work~\cite{imtiaz2025towards}. While the earlier version introduced a
basic SF-UDA pipeline, the current approach incorporates several key
improvements to strengthen target-domain generalization and reduce the
effect of noisy pseudo-labels.

The framework consists of four stages: pre-training, computation,
target adaptation, and inference. Unlike our earlier work, which
relied on supervised source-domain training, we introduce a
non-contrastive predictive coding--based self-supervised strategy to
learn robust and transferable EEG representations from signals. A
predictive coding model is trained to reconstruct future
representations from past activity without using negative pairs or
contrastive learning, using an encoder--decoder architecture optimized
via a predictive reconstruction loss. This design captures temporal
dependencies, encourages domain-invariant features, and removes
reliance on labeled source data after pretraining, making it suitable
for privacy-sensitive settings.

In the computation stage, we estimate the class-wise structure in the
target domain using pretrained predictions to derive cluster centroids
and classifier disagreement for adaptation. The target adaptation
stage consists of two steps using unlabeled data: Multi-Loss Adaptive
Regularization (MLAR), which reduces classifier disagreement on
high-confidence samples while aligning pseudo-labels to stabilize
decision boundaries, and Localized Consistency Learning (LCL), which
enforces prediction consistency among reliable neighbors selected
using both feature- and classifier-space similarity, reducing the
influence of unreliable neighbors.

A key improvement is introduced at inference time. Unlike our previous
PC-TTA method~\cite{imtiaz2025towards}, we introduce a lightweight
test-time training strategy. During inference, samples are evaluated
using prediction entropy and predictive error from the temporal
predictive coding objective. Uncertain samples are selectively updated
with a few gradient steps using predictive and entropy losses,
enabling dynamic adaptation and improving robustness under domain
shift while avoiding unnecessary computation for confident
predictions.

We evaluate the framework on DEAP~\cite{koelstra2011deap},
SEED~\cite{zheng2015investigating}, and
DREAMER~\cite{katsigiannis2017dreamer} in a cross-dataset setting. We
compare against recent SF-UDA
methods~\cite{ragab2023source,zhao2023source,ahmed2023ssda,du2024generation},
and the results show consistent improvements across both binary and
multi-class emotion classification tasks in all cross-dataset
scenarios.

The main contributions of this work are as follows:

\begin{enumerate}
\item We propose an enhanced source-free unsupervised domain
  adaptation framework for cross-dataset EEG-based emotion recognition
  that addresses domain shifts and the absence of source data.

\item We introduce a non-contrastive predictive coding--based
  self-supervised pretraining strategy to learn robust temporal EEG
  representations without relying on negative pairs or contrastive
  objectives.

\item We design a lightweight test-time training mechanism to enhance
  robustness under distribution shift, where uncertain target samples
  are refined online using predictive reconstruction and entropy
  minimization.

\item Extensive experiments demonstrate that the multi-stage framework
  achieves consistent improvements over state-of-the-art SF-UDA
  baselines in cross-dataset EEG emotion recognition, validating the
  combined benefit of domain-invariant representation learning and
  noise-robust pseudo-label optimization.
\end{enumerate}

The remainder of this paper is organized as follows.
Section~\ref{related_work} reviews related work on EEG emotion
recognition, domain adaptation, and source-free learning.
Section~\ref{proposed_method} presents the proposed framework,
including pretraining, adaptation, and test-time training.
Section~\ref{experiments} describes the experimental setup, datasets,
and results. Section~\ref{conclusion} concludes the paper and outlines
future directions.

% -------------------------------------------------------
\section{Related Work}
\label{related_work}

Extensive research has explored EEG-based emotion recognition using
both traditional and deep learning methods. Classical classifiers such
as Support Vector Machines, K-Nearest Neighbors, Decision Trees,
Logistic Regression, and Linear Discriminant
Analysis~\cite{cai2022eeg,patel2024cross,zheng2023adaptive,singh2024emotion,melinda2023classification}
are widely used but struggle with the non-stationary and temporal
nature of EEG signals, limiting generalization to unseen subjects. In
deep learning, models such as Autoencoders, Graph Neural Networks, and
Deep Belief Networks have been
investigated~\cite{pang2024multi,lin2023eeg,wang2023classification},
while Convolutional Neural Networks and Long Short-Term Memory
networks remain popular for capturing spatial--temporal EEG
patterns~\cite{xu2025unsupervised,shi2023enhancing,wang2024eeg,fan2024icaps}.

Deep learning models require large labeled datasets, which are costly
and time-consuming to obtain. Self-supervised learning (SSL) reduces
this dependency by learning from unlabeled data through pretext
tasks~\cite{zhang2024self,ren2025comprehensive}. In EEG, SSL includes
contrastive, autoencoder-based, graph-based, and hybrid
approaches~\cite{wang2023self,montero2022applying}. While contrastive
methods are effective, they depend on negative sampling and large
batch sizes. Predictive coding--based
methods~\cite{sprevak2024predictive,sheibani2026pc} have recently
emerged as a simpler alternative by learning representations through
latent prediction rather than instance discrimination.

Despite these advances, performance often degrades under distribution
shifts across subjects, sessions, and datasets. UDA addresses this by
aligning source and target
distributions~\cite{xu2025unsupervised,zhu2025tensorial}. Existing
EEG-UDA methods focus on feature alignment and pseudo-label
refinement, including fine- and coarse-grained
alignment~\cite{jimenez2023learning} and distribution matching and
confidence filtering~\cite{tang2025uda}. Adversarial approaches
further learn domain-invariant features via competition between
feature extractors and
discriminators~\cite{li2025adversarial,he2022adversarial}.

Only a few
studies~\cite{ni2021domain,gu2022multi,imtiaz2025enhanced} address
cross-dataset EEG emotion recognition, which is particularly
challenging due to demographic variability and recording
inconsistencies. These works explore dictionary
learning~\cite{gu2022multi}, subspace
projection~\cite{ni2021domain}, and test-time
augmentation~\cite{imtiaz2025enhanced} strategies to reduce negative
transfer.

Traditional UDA assumes access to source data during adaptation,
which is often impractical due to privacy, storage, and computational
constraints. Source-Free Domain Adaptation (SFDA) addresses this
issue by transferring knowledge from a pretrained source model to
unlabeled target data. SFDA methods are broadly categorized into
data-centric and model-centric
approaches~\cite{li2024comprehensive}. Data-centric methods enhance
target-domain structure and representation through clustering,
pseudo-source construction, data augmentation, or iterative
pseudo-label
refinement~\cite{liu2025srpl,jiang2025hg,du2024generation,huang2022relative,tian2025source}.
In contrast, model-centric methods directly adapt model on target data
using pseudo-labeling, information maximization, and entropy
minimization to improve decision boundaries and
transferability~\cite{ahmed2023ssda,liang2020we,li2022source,liu2024source,yang2022source}.

Although most SFDA research focuses on computer vision
applications~\cite{du2024generation,ahmed2023ssda,liu2025srpl,jiang2025hg,huang2022relative,yang2022source},
its application to time-series signals such as EEG remains
limited~\cite{ragab2023source,zhao2023source,wang2025active}. Ragab
et al.~\cite{ragab2023source} proposed a framework using random
masking and temporal imputation to align target and source
representations while preserving temporal coherence. Zhao and
Peng~\cite{zhao2023source} introduced SS-TrBoosting, a boosting-based
SFDA method for EEG seizure classification, later extended to an
unsupervised variant (U-TrBoosting). Wang et
al.~\cite{wang2025active} proposed an iEEG-based epilepsy approach
that selects informative samples using neighborhood uncertainty and
diversity for annotation.

Although deep learning has shown strong performance in EEG-based
emotion recognition, cross-domain adaptation remains challenging due
to significant distribution shifts between datasets. This is further
amplified in source-free settings, where source data is unavailable
and conventional alignment methods cannot be applied. Despite its
potential, SFDA has been underexplored for emotion recognition. To
address this gap, we propose a source-free unsupervised domain
adaptation framework for EEG-based emotion classification that handles
cross-dataset variability and the absence of source data during
adaptation.

% -------------------------------------------------------
\section{Proposed Method}
\label{proposed_method}

Our method consists of four stages: pre-training, computation, target
adaptation, and inference. Building on the pipeline of our previous
SF-UDA model~\cite{imtiaz2025towards}, we introduce key modifications
to the pre-training, target adaptation, and inference stages.

\subsection{Framework}

Let the training dataset be the \textit{source domain} and the
testing dataset the \textit{target domain}. The source domain
contains $N_s$ labeled EEG samples
$X_s=\{x_s^i\}_{i=1}^{N_s}$ with labels
$Y_s=\{y_s^i\}_{i=1}^{N_s}$, while the target domain contains $N_t$
unlabeled samples $X_t=\{x_t^i\}_{i=1}^{N_t}$, where $x_s^i$ and
$x_t^i$ are the feature representations of the $i^{th}$ EEG segment.
The two domains follow different marginal distributions,
$P_s(X_s) \neq P_t(X_t)$. We address source-free domain adaptation,
where source data is unavailable during adaptation for privacy
preservation. The objective is to learn a mapping function $f$ that
enables accurate emotion prediction on the target domain by minimizing
domain discrepancy through the adaptation stages described below.

\begin{figure}[!tb]
  \centering
  \includegraphics[width=\linewidth]{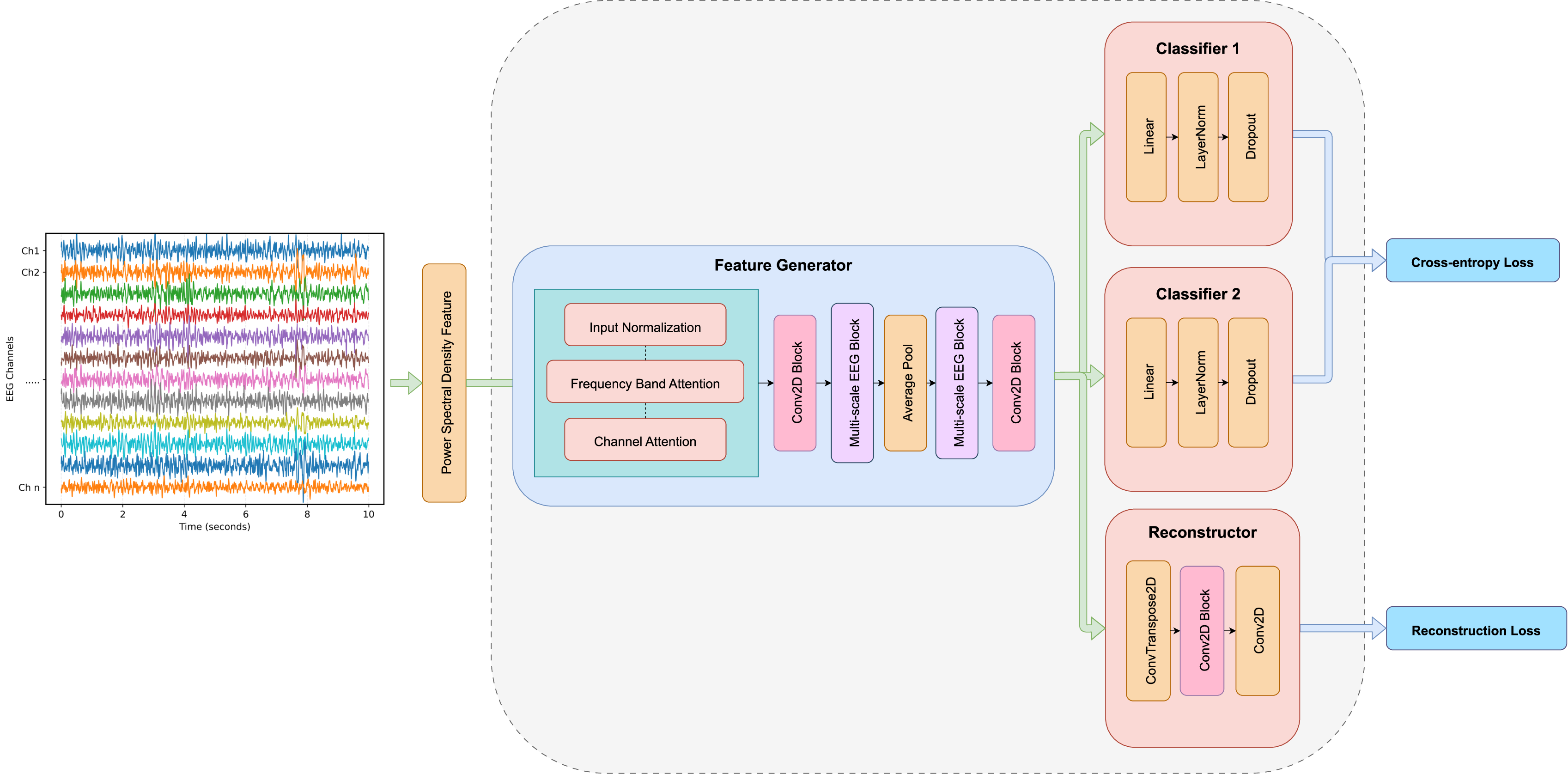}
  \caption{Overall architecture of the proposed network. The feature
    generator integrates learnable input normalization, axis-wise
    attention gating, Conv2D blocks (convolution, group normalization,
    and dropout) for spatial feature learning, and multiscale EEG
    blocks (stacked convolutions, normalization, adaptive pooling, and
    dropout) for multiscale EEG representation learning. A
    reconstructor enables predictive coding via input reconstruction,
    and two parallel classifiers produce the final prediction by
    averaging their outputs.}
  \label{model_framework}
\end{figure}

The proposed network architecture is illustrated in
Fig.~\ref{model_framework}. The feature generator ($F$) applies
learnable input normalization and axis-wise attention gating, followed
by Conv2D and multiscale convolutional blocks with progressive spatial
downsampling to extract hierarchical EEG features. The Conv2D block
comprises convolution, group normalization, and dropout, while the
multiscale EEG block combines stacked convolutions, normalization,
adaptive pooling, and dropout to capture multi-scale representations.

The reconstructor ($R$), used only during pre-training and test-time
training, reconstructs the input via transposed convolutions to
support predictive coding and preserve informative temporal structure.
Two parallel classifiers ($C_1$ and $C_2$), attached to the shared
feature representation as in our previous
works~\cite{imtiaz2025towards,imtiaz2025enhanced,imtiaz2024cross},
generate the final prediction by averaging their outputs. Their
discrepancy is incorporated into the pre-training loss to promote
consistency and robust feature learning. Each classifier consists of
two fully connected layers with Layer Normalization, SiLU activation,
and dropout.

\subsubsection{Pre-training}

During pre-training, the model is optimized on labeled source data
($X_s, Y_s$) to learn discriminative and temporally predictive
representations. The objective includes three terms: classification
loss ($L_{cls}$), classifier discrepancy loss ($L_{dis}$), and
predictive coding loss ($L_{pred}$), as defined in
Eq.~\ref{eq3}. Following~\cite{imtiaz2025towards,imtiaz2025enhanced},
$L_{cls}$ uses weighted cross-entropy, while $L_{dis}$ measures the
Euclidean distance between the two classifier outputs to enforce
consistency.

In addition, motivated by~\cite{sheibani2026pc} and the temporal
continuity of EEG dynamics, this work introduces a predictive coding
module to capture the intrinsic evolution of neural activity.
Specifically, given an input ($x_t$), the feature generator ($F$)
maps it into a latent representation, which is subsequently passed to
a reconstructor ($R$) to predict the power spectral density (PSD)
representation of the next segment:
\begin{align}
  \hat{F}_{t+1} = R(F(x_t))
  \label{eq1}
\end{align}

The predictive coding loss is defined as the mean squared error
between the predicted and ground-truth future PSD features:
\begin{align}
  L_{pred} = \frac{1}{N_{b}} \sum_{i=1}^{N_{b}}
  \left\| \hat{F}_{t+1}^{(i)} - x_{t+1}^{(i)} \right\|_2^2
  \label{eq2}
\end{align}
where $x_{t+1}$ denotes the ground-truth future PSD representation
for the $i^{th}$ sample and $N_{b}$ is the batch size. This objective
encourages the model to capture the underlying temporal structure of
EEG signals, reflecting their continuous and predictive nature.

Finally, the overall pre-training objective is:
\begin{align}
  L = L_{cls} + \alpha_1 L_{dis} + \alpha_2 L_{pred}
  \label{eq3}
\end{align}
where $\alpha_1$ and $\alpha_2$ are weighting factors controlling the
contribution of each loss term. Joint optimization of these objectives
enables the model to learn discriminative, regularized, and temporally
predictive EEG representations.

\subsubsection{Computation}

After pre-training, the trained network is obtained and source data is
no longer accessible. Following~\cite{imtiaz2025towards}, cluster
centroids ($CC_t$) for target adaptation are computed by feeding
unlabeled target data into the pre-trained model. As
in~\cite{imtiaz2025towards,liang2020we}, the centroid for emotion
class $k$ is:
\begin{align}
  CC_t^k = \frac{\Sigma_{x_t\in X_t}\, \hat{p}^k(x_t)F(x_t)}
                {\Sigma_{x_t\in X_t}\, \hat{p}^k(x_t)}
  \label{eq4}
\end{align}
where $\hat{p}^k(x_t)$ denotes the combined softmax probability of
sample $x_t$ produced by the two classifiers for the $k^{th}$ emotion
class:
\begin{align}
  \hat{p}^k(x_t) = avg\!\left(\delta^k(C_1(F(x_t))),\,
                               \delta^k(C_2(F(x_t)))\right)
  \label{eq5}
\end{align}
where $\delta$ represents the softmax function. This approach
computes centroids by combining feature representations with
classifier predictions, weighting each sample by prediction
confidence. In contrast, methods using only feature extractor outputs
produce less stable centroids, as they ignore decision boundaries and
confidence, increasing ambiguity under domain shift.

\subsubsection{Target Adaptation}

The target adaptation stage comprises two training steps that operate
solely on unlabeled target data using a pre-trained network.

In the first step, we introduce a Multi-Loss Adaptive Regularization
(MLAR) framework to enhance consistency and robustness on unlabeled
target-domain data. For each training batch, we first generate
pseudo-labels by assigning each sample to the nearest cluster
centroid in the feature space:
\begin{align}
  \hat{y}_t = \underset{k}{\arg\min}\; D_l(F(x_t), CC_t^k)
  \label{eq6}
\end{align}
where $D_l$ denotes the Euclidean distance computed using the L2
norm. We then construct a reliable training mask by selecting
confident samples ($CX_t$) that satisfy multiple consistency
criteria, including agreement between cluster-based pseudo-labels and
model predictions, and high prediction confidence:
\begin{align}
  CX_t = \bigl\{ x_t \in X_{t,b} \mid
    \hat{y}_t^{(cluster)} = \hat{y}_t^{(model)},\;
    \max(\hat{p}(x_t)) \ge \tau_c \bigr\}
  \label{eq7}
\end{align}
where $X_{t,b}$ denotes target mini-batch samples,
$\hat{y}_t^{(cluster)}$ is the pseudo-label from nearest-centroid
assignment, $\hat{y}_t^{(model)}$ is the classifier prediction,
$\hat{p}(x_t)$ is the combined softmax probability, and $\tau_c$ is
the confidence threshold for selecting reliable samples.

The primary learning objective is the pseudo-label agreement loss
($L_{plal}$), computed over the selected reliable samples:
\begin{align}
  L_{plal} = -\frac{1}{|CX_t|} \sum_{x_t \in CX_t}
  \log \hat{p}_{\hat{y}_t}(x_t)
  \label{eq8}
\end{align}

To further improve classifier consistency, we introduce a classifier
discrepancy loss ($L_{cdl}$), computed as the Jensen--Shannon ($JS$)
divergence between the two classifiers:
\begin{align}
  L_{cdl} = \mathbb{E}_{x_t \in CX_t}
  \Bigl[ JS\bigl(\hat{p}_1(x_t) \big\| \hat{p}_2(x_t)\bigr) \Bigr]
  \label{eq9}
\end{align}
where $\hat{p}_1$ and $\hat{p}_2$ are the softmax outputs of $C_1$
and $C_2$, respectively.

In addition, we apply two auxiliary regularization terms to stabilize
training. Prediction entropy minimization:
\begin{align}
  L_{ent} = -\frac{1}{|CX_t|} \sum_{x_t \in CX_t} \sum_{k=1}^{K}
  \hat{p}^k(x_t)\log \hat{p}^k(x_t)
  \label{eq10}
\end{align}
Target distribution balancing:
\begin{align}
  L_{bal} = \left\| \frac{1}{N_b} \sum_{x_t \in X_{t,b}}
  \hat{p}(x_t) - \frac{1}{K}\mathbf{1}_K \right\|_2^2
  \label{eq11}
\end{align}
where $K$ denotes the number of classes and $\mathbf{1}_K$ is a
$K$-dimensional vector of ones. These terms prevent prediction
collapse and encourage uniform utilization of classes.

Finally, the overall MLAR objective is:
\begin{align}
  L_{mlar} = \lambda_1 L_{plal} + \lambda_2 L_{cdl}
           + \lambda_3 L_{ent} + \lambda_4 L_{bal}
  \label{eq12}
\end{align}
where $\lambda_1$, $\lambda_2$, $\lambda_3$, and $\lambda_4$ are
weighting factors. At the end of each epoch, cluster centroids are
updated using features from the current model, ensuring progressive
alignment between pseudo-labels and the evolving target feature space.

In the second step, we employ the Localized Consistency Learning (LCL)
strategy~\cite{imtiaz2025towards} to enforce local prediction
consistency among neighboring target samples. For each sample,
reliable neighbors are selected using both feature-space and
classifier-output similarities, and the final set is obtained by
intersecting the two nearest-neighbor groups. The model is optimized
using the localized consistency loss ($L_{lcl}$), which minimizes
prediction discrepancies among these neighbors. The neighbor selection
and loss formulation follow~\cite{imtiaz2025towards}.

MLAR improves prediction reliability on unlabeled target data by
reducing noise and enforcing consistency. LCL further enhances local
agreement among reliable neighbors by aligning feature and classifier
spaces.

\subsubsection{Inference}

\begin{figure}[!tb]
  \centering
  \includegraphics[width=\linewidth]{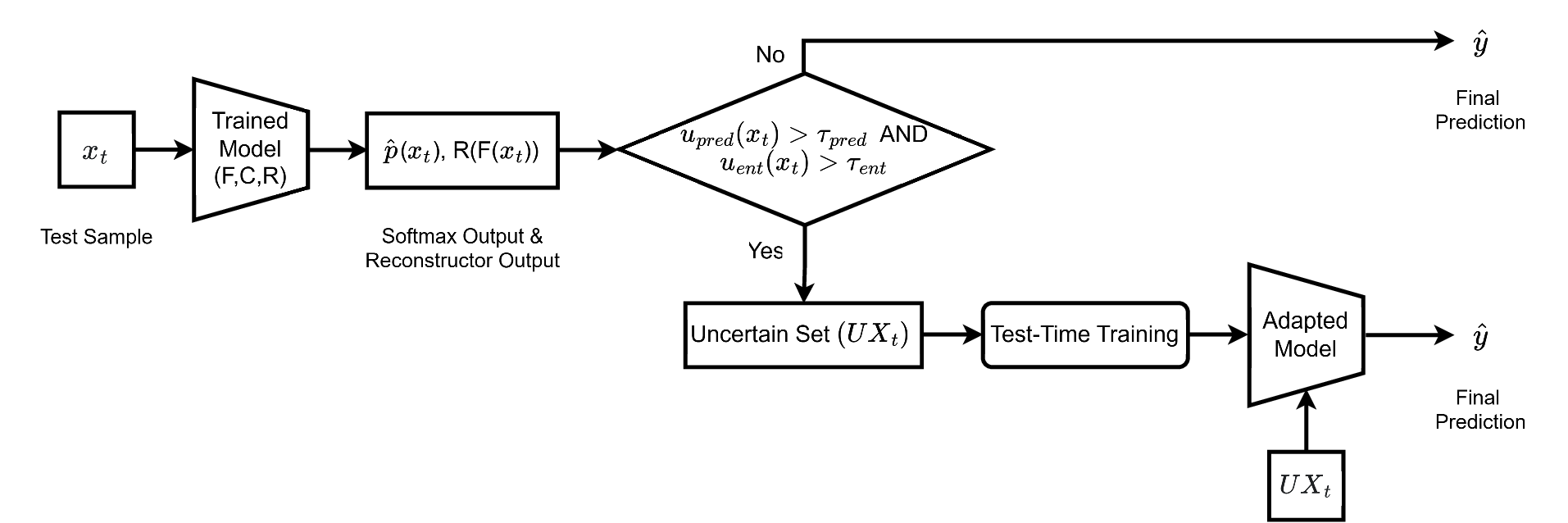}
  \caption{Proposed uncertainty-based test-time training process.}
  \label{tttFig}
\end{figure}

We propose uncertainty-based test-time training
(Fig.~\ref{tttFig}) on target samples during inference to mitigate
incorrect predictions while improving computational efficiency. During
test-time training, the model is updated online using unlabeled target
data. To ensure stability, unreliable samples are identified based on
predictive uncertainty. For each $x_t$, we compute reconstruction
error and prediction entropy:
\begin{align}
  u_{pred}(x_t) &= \| R(F(x_t)) - x_t \|_2^2  \label{eq13} \\
  u_{ent}(x_t) &= -\sum_{k=1}^{K} \hat{p}^k(x_t)\log \hat{p}^k(x_t)
  \label{eq14}
\end{align}

Thresholds are defined using the quantile statistics of the current
batch:
\begin{align}
  \tau_{pred} &= \mathrm{Quantile}\bigl(u_{pred}(X_t), q\bigr),\quad
  \tau_{ent}  = \mathrm{Quantile}\bigl(u_{ent}(X_t), q\bigr)
  \label{eq15}
\end{align}
where $q=0.5$ in our implementation. Based on these thresholds, we
construct the uncertain target subset $UX_t$:
\begin{align}
  UX_t = \left\{ x_t \in X_t \;\middle|\;
    u_{pred}(x_t) > \tau_{pred} \;\land\;
    u_{ent}(x_t) > \tau_{ent}
  \right\}
  \label{eq16}
\end{align}

Only selected samples $UX_t$ are used for test-time adaptation,
reducing the impact of incorrect predictions and improving
computational efficiency. Samples that do not satisfy the uncertainty
criterion are excluded, and their predictions remain unchanged. On
$UX_t$, only lightweight components (normalization layers and
attention gate modules) are updated using entropy and
reconstruction-based objectives. Adaptation is performed for a
limited number of iterations to ensure computational efficiency.

Entropy minimization loss:
\begin{align}
  L_{ttent} = -\frac{1}{|UX_t|} \sum_{x_t \in UX_t} \sum_{k=1}^{K}
  \hat{p}^k(x_t)\log \hat{p}^k(x_t)
  \label{eq17}
\end{align}

Predictive reconstruction loss (unlike Eq.~\ref{eq2}, which predicts
the \emph{next} segment's PSD during pre-training, this loss
reconstructs the \emph{current} sample to measure uncertainty at test
time):
\begin{align}
  L_{ttpred} = \frac{1}{|UX_t|} \sum_{x_t \in UX_t}
  \left\| R(F(x_t)) - x_t \right\|_2^2
  \label{eq18}
\end{align}

Finally, the overall test-time training objective is:
\begin{align}
  L_{ttt} = \beta_1 L_{ttent} + \beta_2 L_{ttpred}
  \label{eq19}
\end{align}
where $\beta_1$ and $\beta_2$ are weights for each loss term.
Following test-time training with objective $L_{ttt}$, final
predictions for uncertain samples are obtained using the adapted
model. Algorithm~\ref{alg} summarizes the complete procedure.

\begin{algorithm}[!tb]
\footnotesize
\begin{algorithmic}
\REQUIRE Source PSD samples $X_s$, Target PSD samples $X_t$,
         Source labels $Y_s$; Feature generator $F$, classifiers
         $C_1$, $C_2$, reconstructor $R$;
         Epochs: $Epoch_{mlar}$, $Epoch_{lcl}$

\ENSURE
\textbf{Pre-training:}
Train on $(X_s, Y_s)$: minimize $L$ (Eq.~\ref{eq3}), comprising
$L_{cls}$, $L_{dis}$, and $L_{pred}$ (Eq.~\ref{eq2})

\textbf{Computation:}
Compute cluster centroids $CC_t$ (Eq.~\ref{eq4})

\textbf{Target Adaptation:}
\For{$i = 1$ to $Epoch_{mlar}$}{
  Generate pseudo-labels (Eq.~\ref{eq6});
  construct $CX_t$ (Eq.~\ref{eq7})\\
  Minimize $L_{mlar}$ (Eq.~\ref{eq12}) on $X_t$
}
\For{$i = 1$ to $Epoch_{lcl}$}{
  Identify reliable neighbors; minimize $L_{lcl}$ on $X_t$
}

\textbf{Inference:}
\For{each mini-batch $X_t$}{
  Compute $u_{pred}$ (Eq.~\ref{eq13}),
          $u_{ent}$ (Eq.~\ref{eq14}),
          thresholds (Eq.~\ref{eq15}),
          $UX_t$ (Eq.~\ref{eq16})\\
  Predict certain samples ($X_t \setminus UX_t$) without adaptation\\
  \If{$UX_t \neq \emptyset$}{
    Update normalization layers and attention gates by minimizing
    $L_{ttt}$ (Eq.~\ref{eq19}) on $UX_t$\\
    Predict samples in $UX_t$ using the adapted model
  }
}
\end{algorithmic}
\caption{Proposed SF-UDA approach for EEG emotion recognition.}
\label{alg}
\end{algorithm}

\subsection{Data Preprocessing and Model Input Construction}

We use Power Spectral Density (PSD) and Differential Entropy (DE) as
EEG features, consistent with prior
studies~\cite{ni2021domain,li2022eeg}. Although both are evaluated,
PSD consistently shows better performance and is therefore adopted in
this study. The preprocessing is the same as in our previous
methods~\cite{imtiaz2025towards,imtiaz2025enhanced}. Signals are
extracted from 32 common channels shared by the DEAP and SEED datasets
to ensure consistent input dimensions. Each trial is segmented into
2-second windows with a 1-second overlap, and PSD features are
computed across delta (1--3~Hz), theta (4--7~Hz), alpha (8--13~Hz),
beta (14--30~Hz), and gamma (31--50~Hz) bands. The resulting features
are organized into a 2-D matrix $X \in \mathbb{R}^{n \times i}$,
where $n=32$ and $i=5$, yielding a $32 \times 5$ representation. For
the DREAMER dataset, 14 channels are used, producing a $14 \times 5$
representation. Finally, all features are normalized to the range
$[-1, 1]$.

% -------------------------------------------------------
\section{Experiments}
\label{experiments}

\subsection{Experimental Details}

We evaluate the proposed model on the publicly available DEAP, SEED,
and DREAMER datasets, which are widely used in emotion recognition.
Cross-dataset experiments are performed by training on one dataset and
testing on another.

\paragraph{DEAP~\cite{koelstra2011deap}.}
The DEAP dataset involves 32 participants who viewed 40 one-minute
music videos. Each trial lasted 63 seconds, including a 3-second
pre-trial interval followed by a 60-second video presentation. After
each trial, participants rated arousal, valence, dominance, and liking
on a 1--9 scale. EEG was recorded using 32 electrodes at 512~Hz, then
downsampled to 128~Hz and filtered with a 4--45~Hz band-pass filter.

\paragraph{SEED~\cite{zheng2015investigating}.}
The SEED dataset consists of EEG recordings from 15 participants
across three sessions held on different days. In each session,
participants viewed 15 Chinese film clips designed to elicit positive,
neutral, and negative emotions. EEG signals were recorded using
62-channel electrodes at 1,000~Hz, then down-sampled to 200~Hz and
band-pass filtered between 0--75~Hz.

\paragraph{DREAMER~\cite{katsigiannis2017dreamer}.}
The DREAMER dataset contains EEG recordings from 23 participants who
watched 18 film clips of varying durations (65--393~seconds). After
each clip, participants rated valence, arousal, and dominance on a
1--5 scale. EEG data were collected using 14 electrodes at 128~Hz.

We follow the preprocessing protocol of our previous
studies~\cite{imtiaz2025towards,imtiaz2025enhanced}. All subjects and
trials are used. In DEAP, the first 3 seconds of each trial are
removed. SEED and DREAMER signals are band-pass filtered between
0.3--50~Hz~\cite{li2019domain,imtiaz2025towards,imtiaz2025enhanced},
while DEAP is kept unchanged as it is already filtered (4--45~Hz). We
perform both binary and three-class emotion recognition. For binary
classification, DEAP uses a valence threshold of 4.5, DREAMER uses
3.0~\cite{ni2021domain,zhang2024tpro,imtiaz2025towards}, and SEED
includes only positive and negative classes. For multi-class
classification, DEAP is split into negative (1--4), neutral (4--6),
and positive (6--9)~\cite{tripathi2017using,imtiaz2025towards}, SEED
retains its original labels, and DREAMER is labeled as negative
(1--2), neutral (3), and positive (4--5).

All experiments are conducted on a Linux system using Python 3.12.13
and PyTorch 2.11.0+cu128, with an NVIDIA Tesla T4 GPU (15~GB). The
batch size is 64. The hyperparameters $\alpha_1$, $\alpha_2$,
$\lambda_1$, $\lambda_2$, $\lambda_3$, $\lambda_4$, $\beta_1$,
$\beta_2$, and $\tau_c$ are set to 0.5, 0.5, 0.5, 0.5, 0.1, 0.1,
0.1, 0.5, and 0.8 respectively.

\subsection{Results and Discussion}

We evaluate performance using cross-dataset testing by training on one
dataset and testing on another in both directions. Our method is
compared with our previous SF-UDA approach~\cite{imtiaz2025towards}.
Since no other SF-UDA methods have been specifically designed for EEG
emotion recognition, we further compare against four general-purpose
state-of-the-art SF-UDA
methods~\cite{ragab2023source,zhao2023source,ahmed2023ssda,du2024generation}.
All open-source baselines are reimplemented and evaluated on the same
datasets to ensure a fair comparison.

\begin{table}[!tb]
\scriptsize
\centering
\caption{Binary emotion classification accuracy comparison with
state-of-the-art SF-UDA methods. Statistical significance is
indicated using a paired-sample \textit{t}-test: $\sim$
\textit{nonsignificant}, *$p < 0.05$, **$p < 0.01$.}
\begin{tabular}{lcc}
\toprule
 & \multicolumn{2}{c}{Accuracy (\%)} \\
\cmidrule{2-3}
Method & SEED$\rightarrow$DEAP & DEAP$\rightarrow$SEED \\
\midrule
Ahmed et al.~\cite{ahmed2023ssda}              & 51.71** & 61.33** \\
Du et al.~\cite{du2024generation}              & 48.74** & 56.45** \\
Ragab et al.~\cite{ragab2023source}            & 53.55** & 59.57** \\
Zhao et al.~\cite{zhao2023source}              & 53.17** & 61.98** \\
Our previous SF-UDA~\cite{imtiaz2025towards}   & 58.99*  & 65.84*  \\
\textbf{Proposed method}                       & \textbf{61.38} & \textbf{69.56} \\
\bottomrule
\end{tabular}
\label{table1}
\end{table}

Table~\ref{table1} presents the binary cross-dataset emotion
classification results. The proposed method consistently achieves the
best performance in both transfer directions, reaching 61.38\% for
SEED $\rightarrow$ DEAP and 69.56\% for DEAP $\rightarrow$ SEED,
outperforming all competing SF-UDA methods, including our previous
approach~\cite{imtiaz2025towards}. Paired-sample \textit{t}-tests
confirm that improvements are statistically significant (*) compared
with our previous method and highly significant (**) compared with the
remaining baselines.

\begin{table}[!tb]
\scriptsize
\centering
\caption{Multiclass emotion (positive, neutral, negative)
classification accuracy. Significance notation same as
Table~\ref{table1}.}
\begin{tabular}{lcc}
\toprule
 & \multicolumn{2}{c}{Accuracy (\%)} \\
\cmidrule{2-3}
Method & SEED$\rightarrow$DEAP & DEAP$\rightarrow$SEED \\
\midrule
Ahmed et al.~\cite{ahmed2023ssda}              & 44.08** & 51.29** \\
Du et al.~\cite{du2024generation}              & 39.93** & 47.35** \\
Ragab et al.~\cite{ragab2023source}            & 45.97** & 51.82** \\
Zhao et al.~\cite{zhao2023source}              & 47.45** & 53.60*  \\
Our previous SF-UDA~\cite{imtiaz2025towards}   & 51.50*  & \textbf{57.37} \\
\textbf{Proposed method}                       & \textbf{53.87} & 57.08 \\
\bottomrule
\end{tabular}
\label{table1_b}
\end{table}

We further evaluate the model on the three-class emotion recognition
task (Table~\ref{table1_b}). The proposed method achieves
significantly better performance on SEED $\rightarrow$ DEAP compared
to all competing approaches. For DEAP $\rightarrow$ SEED, our
previous SF-UDA method~\cite{imtiaz2025towards} performs slightly
better, while the proposed method remains highly competitive.

\begin{table*}[!tb]
\scriptsize
\centering
\caption{Emotion classification accuracy comparison with other
SF-UDA methods for binary and three-class settings, where DREAMER
is used as the target dataset. Significance notation same as
Table~\ref{table1}.}
\begin{tabular}{lcccc}
\toprule
 & \multicolumn{2}{c}{Binary Emotion (\%)}
 & \multicolumn{2}{c}{Three-Class Emotion (\%)} \\
\cmidrule(lr){2-3}\cmidrule(lr){4-5}
Method & SEED$\rightarrow$DREAMER & DEAP$\rightarrow$DREAMER
       & SEED$\rightarrow$DREAMER & DEAP$\rightarrow$DREAMER \\
\midrule
Ahmed et al.~\cite{ahmed2023ssda}            & 51.47** & 44.08** & 46.39** & 37.28** \\
Du et al.~\cite{du2024generation}            & 55.65** & 54.46** & 48.65** & 49.00*  \\
Ragab et al.~\cite{ragab2023source}          & 58.26** & 54.90** & 49.02** & 49.15*  \\
Zhao et al.~\cite{zhao2023source}            & 60.29** & 53.75** & 55.96** & 50.80*  \\
Our previous SF-UDA~\cite{imtiaz2025towards} & 67.08   & 58.87** & 59.44*  & 53.83*  \\
\textbf{Proposed method}                     & \textbf{68.90} & \textbf{63.03}
                                             & \textbf{62.67} & \textbf{56.24} \\
\bottomrule
\end{tabular}
\label{table1_c}
\end{table*}

To further assess generalization, additional experiments are conducted
using DREAMER as the target dataset. As shown in
Table~\ref{table1_c}, the proposed method achieves the best
performance across all adaptation settings. For binary classification,
it attains 68.90\% for SEED$\rightarrow$DREAMER and 63.03\% for
DEAP$\rightarrow$DREAMER, surpassing our previous SF-UDA method by
1.82\% and 4.16\%, respectively. In the three-class setting, it also
achieves the highest accuracies of 62.67\% and 56.24\%.

\begin{figure}[!tb]
  \centering
  \subfloat[]{\includegraphics[width=0.48\linewidth]{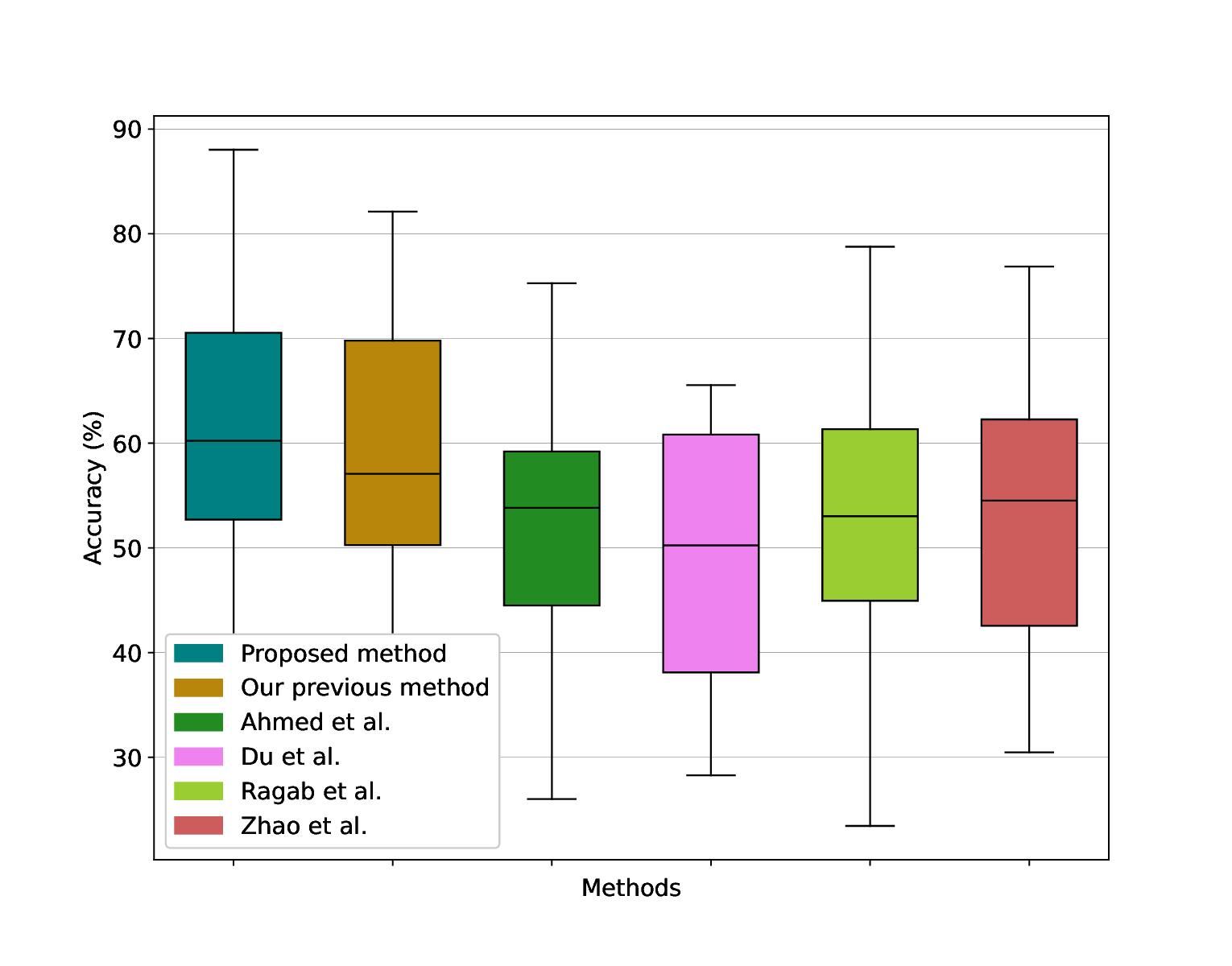}}\hfill
  \subfloat[]{\includegraphics[width=0.48\linewidth]{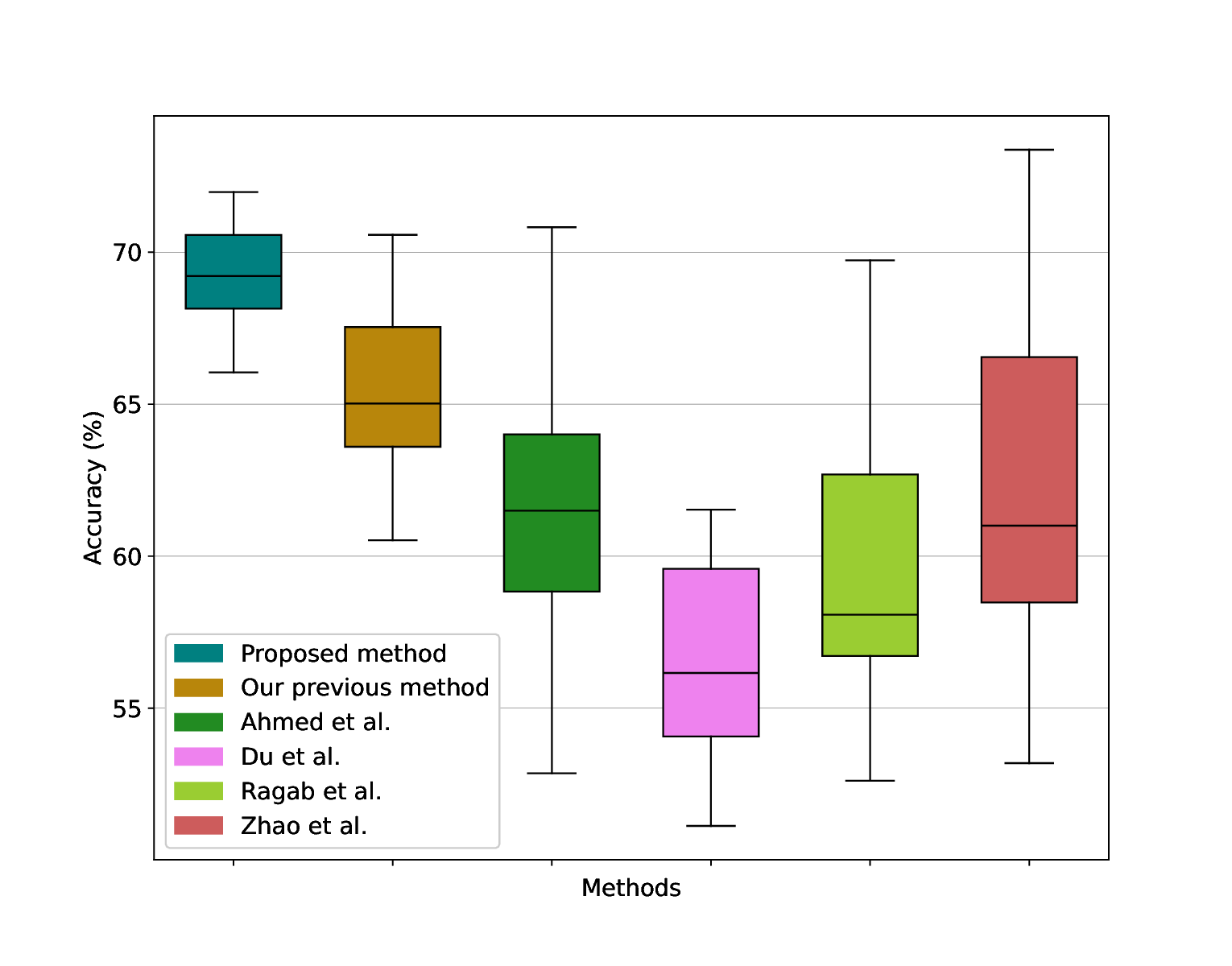}}
  \caption{Subject-level accuracy distributions using boxplots under
  cross-dataset conditions: (a) SEED $\rightarrow$ DEAP and (b) DEAP
  $\rightarrow$ SEED. The proposed model achieves the highest median
  accuracy in both cases while exhibiting lower variability across
  subjects.}
  \label{boxplot}
\end{figure}

Fig.~\ref{boxplot} shows boxplots of subject-wise accuracy
distributions. For DEAP$\rightarrow$SEED, it achieves a median
accuracy of 69.21\% with noticeably lower variability than other
methods. In SEED$\rightarrow$DEAP, it obtains a median accuracy of
60.22\%, again outperforming all baselines. Overall, the median
accuracy improves by 4.19\% and 3.15\% for DEAP$\rightarrow$SEED and
SEED$\rightarrow$DEAP, respectively, over our previous
method~\cite{imtiaz2025towards}.

\begin{table*}[!tb]
\scriptsize
\centering
\caption{Performance comparison (PPV, sensitivity, and F1) for
positive and negative emotion identification.}
\begin{tabular}{lcccccccccccc}
\toprule
 & \multicolumn{6}{c}{SEED$\rightarrow$DEAP}
 & \multicolumn{6}{c}{DEAP$\rightarrow$SEED} \\
\cmidrule(lr){2-7}\cmidrule(lr){8-13}
 & \multicolumn{3}{c}{Negative} & \multicolumn{3}{c}{Positive}
 & \multicolumn{3}{c}{Negative} & \multicolumn{3}{c}{Positive} \\
\cmidrule(lr){2-4}\cmidrule(lr){5-7}\cmidrule(lr){8-10}\cmidrule(lr){11-13}
Method & PPV & Se & F1 & PPV & Se & F1 & PPV & Se & F1 & PPV & Se & F1 \\
\midrule
Ahmed et al.~\cite{ahmed2023ssda}
  & 32.67 & 29.11 & 30.79 & 61.04 & 64.93 & 62.92
  & 66.44 & 45.73 & 54.17 & 58.65 & 76.91 & 66.55 \\
Du et al.~\cite{du2024generation}
  & 26.72 & 22.35 & 24.34 & 58.57 & 64.17 & 61.24
  & 59.00 & 42.18 & 49.19 & 55.03 & 70.70 & 61.89 \\
Ragab et al.~\cite{ragab2023source}
  & 33.38 & 26.01 & 29.24 & 61.69 & 69.66 & 65.43
  & 62.15 & 48.87 & 54.72 & 57.90 & 70.26 & 63.48 \\
Zhao et al.~\cite{zhao2023source}
  & 35.31 & 32.36 & 33.77 & 62.30 & 65.34 & 63.78
  & 64.93 & 52.04 & 57.77 & 60.01 & 71.91 & 65.42 \\
Our previous~\cite{imtiaz2025towards}
  & 41.59 & 27.57 & 33.16 & 64.63 & 77.37 & 70.43
  & 72.70 & 50.69 & 59.73 & 62.17 & 80.97 & 70.34 \\
\textbf{Proposed}
  & 46.53 & 31.40 & \textbf{37.50} & 66.30 & 78.91 & \textbf{72.06}
  & 71.65 & 64.71 & \textbf{68.00} & 67.85 & 74.41 & \textbf{70.98} \\
\bottomrule
\end{tabular}
\label{table2}
\end{table*}

\begin{figure}[!tb]
  \centering
  \includegraphics[width=\linewidth]{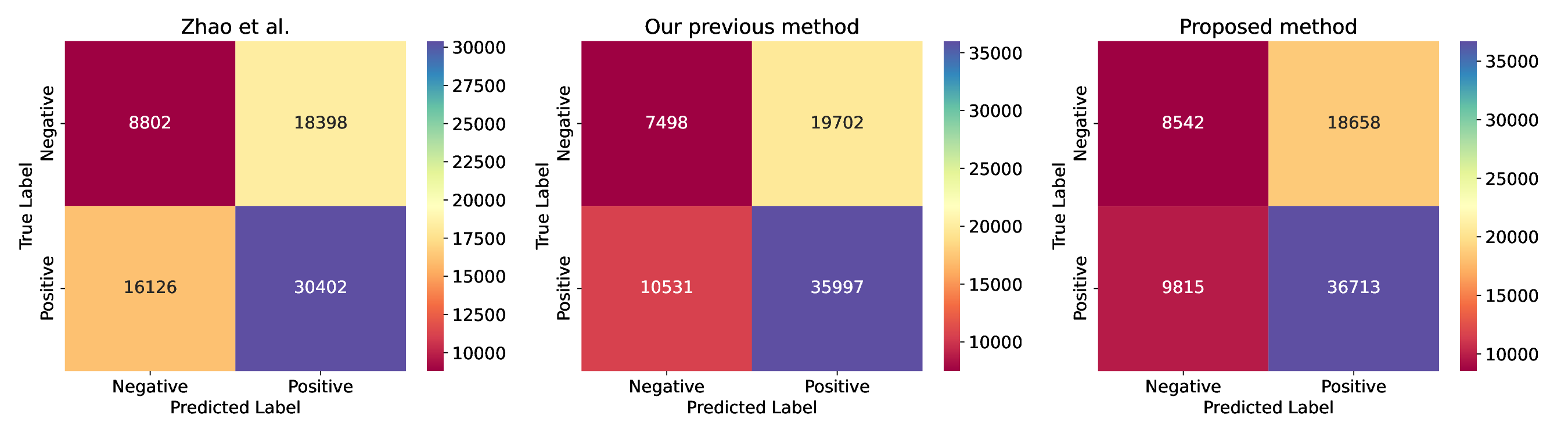}
  \caption{Confusion matrices comparing the proposed method with the
    strongest SF-UDA baselines on SEED $\rightarrow$ DEAP,
    demonstrating improved class-wise consistency and reduced
    inter-class confusion.}
  \label{CFmatix}
\end{figure}

Table~\ref{table2} and Fig.~\ref{CFmatix} compare the performance of
the proposed method with existing SF-UDA approaches for binary emotion
recognition. The proposed framework achieves the highest F1-scores
across both transfer settings, reaching 37.50\% (negative) and
72.06\% (positive) for SEED$\rightarrow$DEAP, and 68.00\% (negative)
and 70.98\% (positive) for DEAP$\rightarrow$SEED. The confusion
matrices further compare the proposed method with the best-performing
competing SF-UDA approaches, demonstrating fewer misclassifications
and more consistent class-wise predictions. The lower performance on
DEAP may be partly attributable to its class imbalance, whereas SEED
is more balanced.

\begin{figure}[!tb]
  \centering
  \subfloat[]{\includegraphics[width=0.48\linewidth]{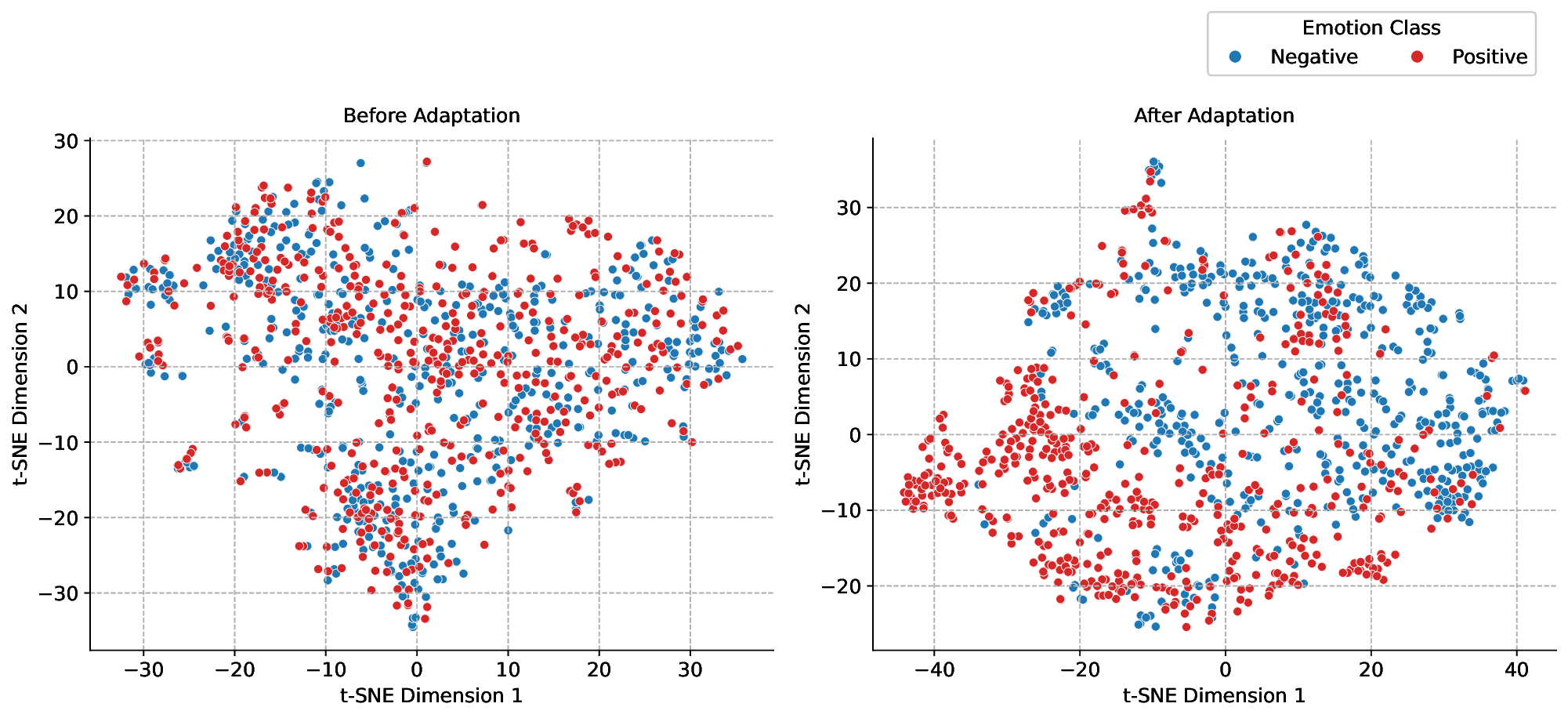}}\\
  \subfloat[]{\includegraphics[width=0.48\linewidth]{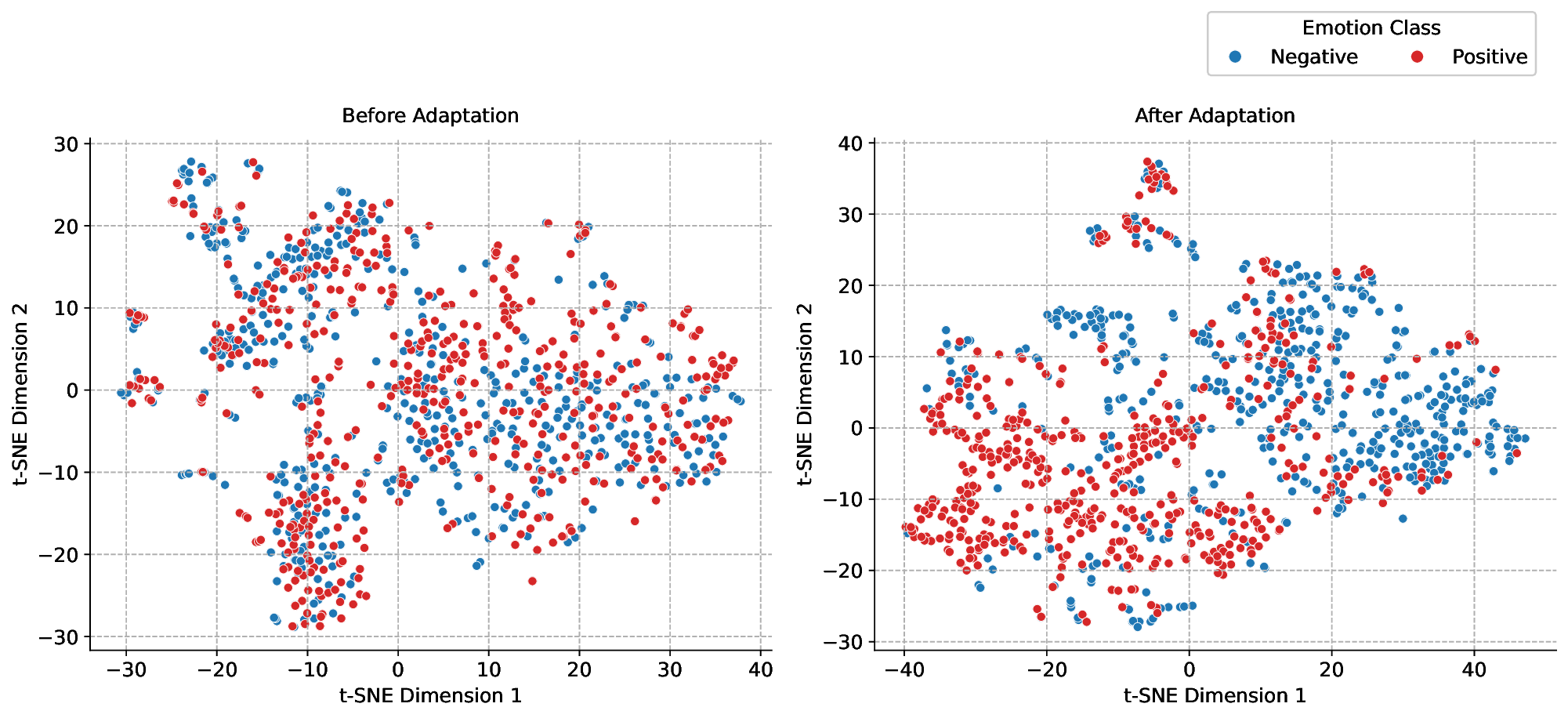}}
  \caption{t-SNE visualization of feature distributions before and
    after adaptation: (a) SEED $\rightarrow$ DEAP and (b) DEAP
    $\rightarrow$ SEED. The proposed method produces more
    discriminative and well-separated embeddings after adaptation.}
  \label{tSNE}
\end{figure}

To qualitatively evaluate the adaptation process, we visualize the
learned feature representations using t-SNE for the
SEED$\rightarrow$DEAP and DEAP$\rightarrow$SEED settings
(Fig.~\ref{tSNE}). Before adaptation, features from different emotion
classes exhibit substantial overlap. After adaptation, the features
become more compact and better separated. Although perfect separation
is not achieved, the improved clustering pattern indicates enhanced
feature alignment and class discriminability.

\subsubsection{Ablation study}

We perform an ablation analysis to investigate the effectiveness of
the major components. Three ablated variants are considered: Model A
excludes the complete adaptation process; Model B removes the MLAR
module; Model C disables test-time training during inference while
preserving all other components.

\begin{table}[!tb]
\scriptsize
\centering
\caption{Overall accuracy comparison in the ablation study (PSD and
DE features).}
\begin{tabular}{lcccc}
\toprule
 & \multicolumn{2}{c}{SEED$\rightarrow$DEAP (\%)}
 & \multicolumn{2}{c}{DEAP$\rightarrow$SEED (\%)} \\
\cmidrule(lr){2-3}\cmidrule(lr){4-5}
Model & PSD & DE & PSD & DE \\
\midrule
Model A & 51.28 & 51.11 & 57.00 & 56.80 \\
Model B & 57.09 & 55.34 & 63.96 & 62.49 \\
Model C & 57.33 & 57.27 & 64.15 & 64.68 \\
\textbf{Proposed} & \textbf{61.38} & 59.93 & \textbf{69.56} & 69.14 \\
\bottomrule
\end{tabular}
\label{table3}
\end{table}

In addition to evaluating models using PSD features, we further
examine their performance with DE features (Table~\ref{table3}).
Overall, PSD achieves better performance, except for a slight
advantage of DE for Model C in the DEAP $\rightarrow$ SEED setting.
PSD is therefore adopted for the final framework. Model A consistently
yields the lowest performance, highlighting the importance of
target-domain adaptation. Removing MLAR (Model B) reduces accuracy by
4.29\% and 5.60\% for SEED $\rightarrow$ DEAP and DEAP $\rightarrow$
SEED, respectively (PSD). Disabling test-time training (Model C)
decreases accuracy by 4.05\% and 5.41\%. These findings confirm the
significant contributions of both modules.

\begin{table}[!tb]
\scriptsize
\centering
\caption{Effect of different uncertainty-based gating strategies in
test-time training.}
\begin{tabular}{lcc}
\toprule
 & \multicolumn{2}{c}{Accuracy (\%)} \\
\cmidrule{2-3}
Uncertainty selection strategy & SEED$\rightarrow$DEAP & DEAP$\rightarrow$SEED \\
\midrule
Entropy + Discrepancy         & 59.74 & 68.10 \\
Entropy + Predictive error    & \textbf{61.38} & \textbf{69.56} \\
Discrepancy + Predictive error & 60.44 & 68.88 \\
All three criteria combined   & 60.35 & 69.27 \\
\bottomrule
\end{tabular}
\label{table3_b}
\end{table}

The results in Table~\ref{table3_b} illustrate the impact of
different uncertainty-based gating strategies within the TTT
framework. Combining entropy with predictive error yields the most
consistent and best performance, suggesting that predictive error
provides more reliable sample-level guidance than discrepancy when
paired with entropy.

\begin{table}[!tb]
\scriptsize
\centering
\caption{Ablation studies of Predictive Coding (PredCod) integration
and test-time training (TTT) components.}

\textbf{(a) PredCod integration strategy}

\begin{tabular}{lcc}
\toprule
 & \multicolumn{2}{c}{Accuracy (\%)} \\
\cmidrule{2-3}
PredCod integration strategy & SEED$\rightarrow$DEAP & DEAP$\rightarrow$SEED \\
\midrule
No PredCod                                         & 56.30 & 62.37 \\
PredCod in pre-training only                       & 59.68 & 67.41 \\
PredCod in TTT only                                & 57.39 & 66.06 \\
PredCod in both (\textbf{proposed})                & \textbf{61.38} & \textbf{69.56} \\
\bottomrule
\end{tabular}

\vspace{2mm}

\textbf{(b) Model update components in TTT}

\begin{tabular}{lcccc}
\toprule
 & \multicolumn{2}{c}{SEED$\rightarrow$DEAP}
 & \multicolumn{2}{c}{DEAP$\rightarrow$SEED} \\
\cmidrule(lr){2-3}\cmidrule(lr){4-5}
TTT update components & Acc.(\%) & Time(s) & Acc.(\%) & Time(s) \\
\midrule
No adaptation                          & 57.33 &  7.62 & 64.15 & 11.00 \\
Classifier only                        & 59.05 & 24.88 & 65.80 & 34.51 \\
Normalization layers only              & 60.21 & 18.11 & 68.11 & 27.53 \\
Attention modules only                 & 58.84 & 20.36 & 67.47 & 30.04 \\
Full model updates                     & 60.13 & 47.40 & 69.73 & 64.72 \\
Norm.\ + Attention (\textbf{proposed}) & \textbf{61.38} & 22.04 & \textbf{69.56} & 31.28 \\
\bottomrule
\end{tabular}
\label{table:ablation_combined}
\end{table}

Table~\ref{table:ablation_combined}(a) presents an ablation study on
the impact of Predictive Coding (PredCod) during pre-training and
TTT. PredCod consistently improves performance in both settings. The
best results are obtained when PredCod is used in both stages,
highlighting the benefit of modeling intrinsic neural dynamics
throughout the learning pipeline.

Table~\ref{table:ablation_combined}(b) evaluates different TTT
component update strategies. All TTT variants outperform the
non-adaptive baseline, confirming the effectiveness of test-time
adaptation. Among partial updates, normalization layers yield more
stable gains. Although full model updates slightly outperform our
method for DEAP$\rightarrow$SEED (by 0.17\%), they underperform for
SEED$\rightarrow$DEAP, primarily due to overfitting on uncertain and
potentially noisy test samples. Full model updates are also the most
computationally expensive, requiring 2.15$\times$ and 2.07$\times$
more time than the proposed method for SEED$\rightarrow$DEAP and
DEAP$\rightarrow$SEED, respectively. The proposed selective adaptation
achieves a better trade-off between efficiency and performance,
reducing inference overhead compared to
PC-TTA~\cite{imtiaz2025towards,imtiaz2025enhanced} by avoiding
multiple augmented forward passes and using fewer update steps.

\subsubsection{EEG feature significance and interpretability analysis}

\begin{figure}[!tb]
  \centering
  \subfloat[]{\includegraphics[width=0.48\linewidth]{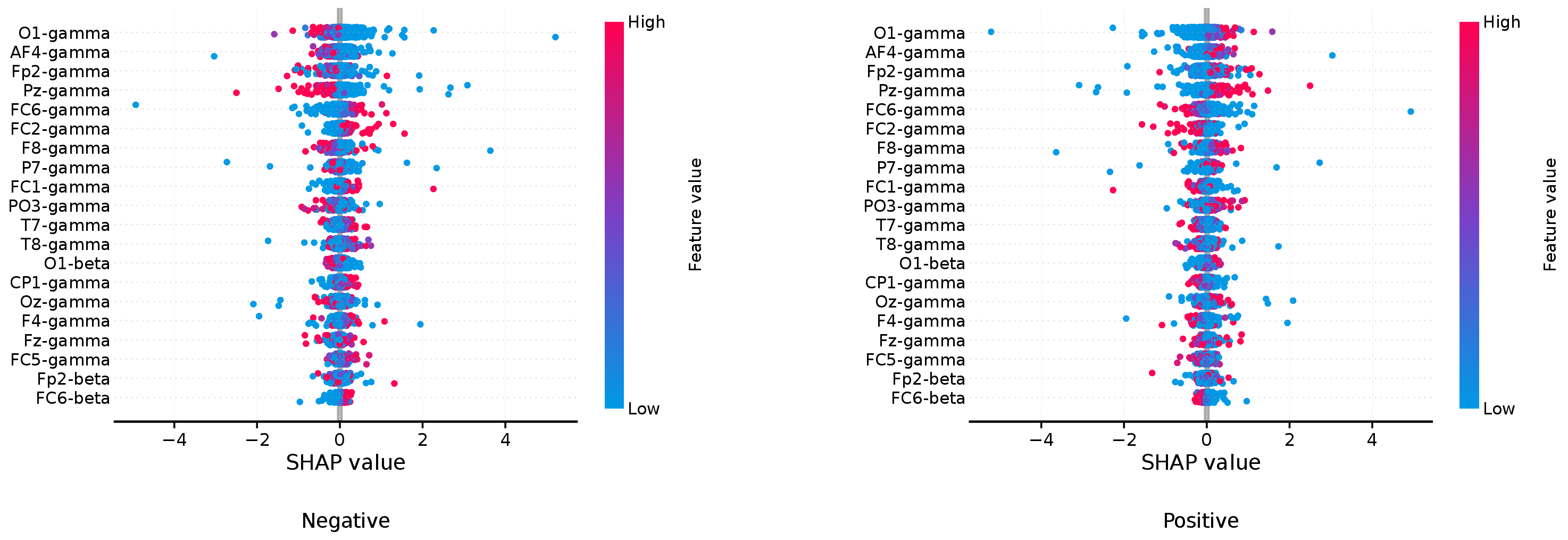}}\hfill
  \subfloat[]{\includegraphics[width=0.48\linewidth]{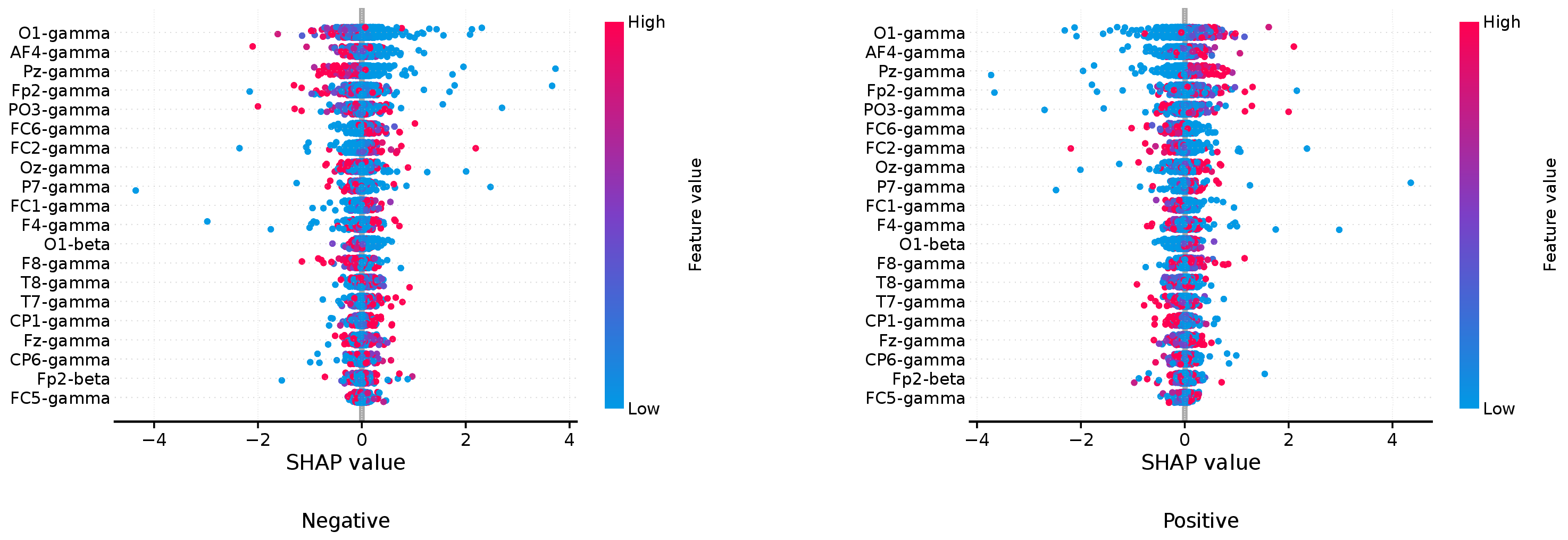}}
  \caption{Distributions of SHAP values for representative subjects
    under cross-dataset settings: (a) SEED $\rightarrow$ DEAP
    (Subject 3), and (b) DEAP $\rightarrow$ SEED (Subject 1). SHAP
    analysis is performed for positive and negative emotion classes,
    showing the top 20 EEG features ranked by importance. Features
    are denoted as channel--frequency band pairs (e.g., O1--gamma).
    Wider dispersion reflects stronger feature influence; dot color
    encodes feature magnitude from low (blue) to high (red).}
  \label{SHAP}
\end{figure}

\begin{figure}[!tb]
  \centering
  \subfloat[]{\includegraphics[width=0.32\linewidth]{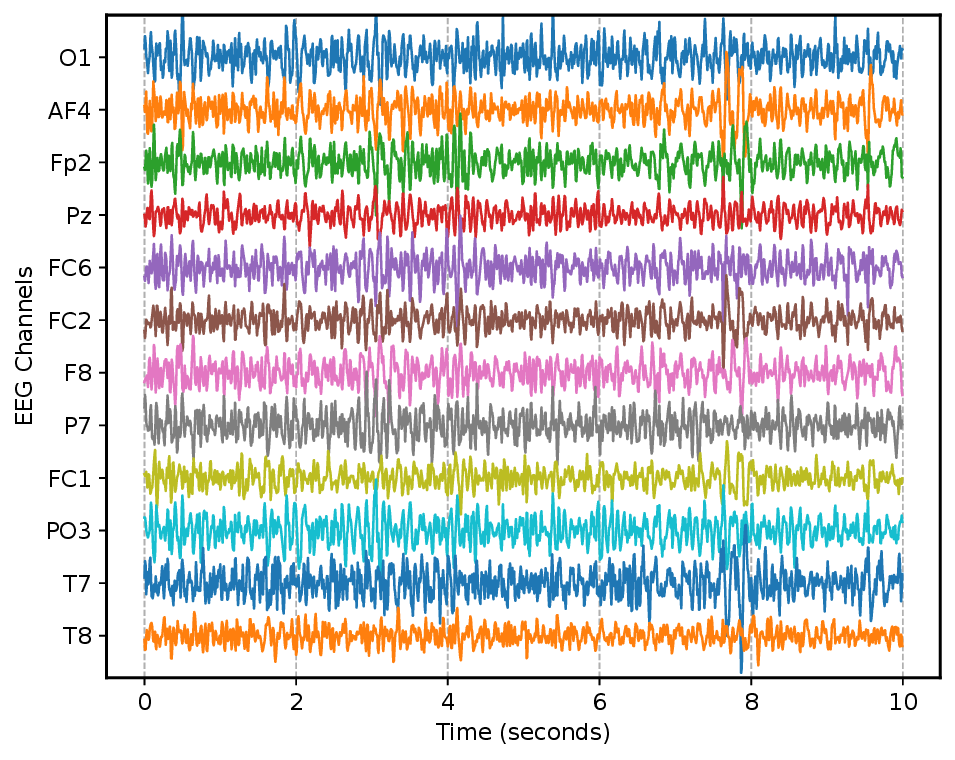}}\hfill
  \subfloat[]{\includegraphics[width=0.32\linewidth]{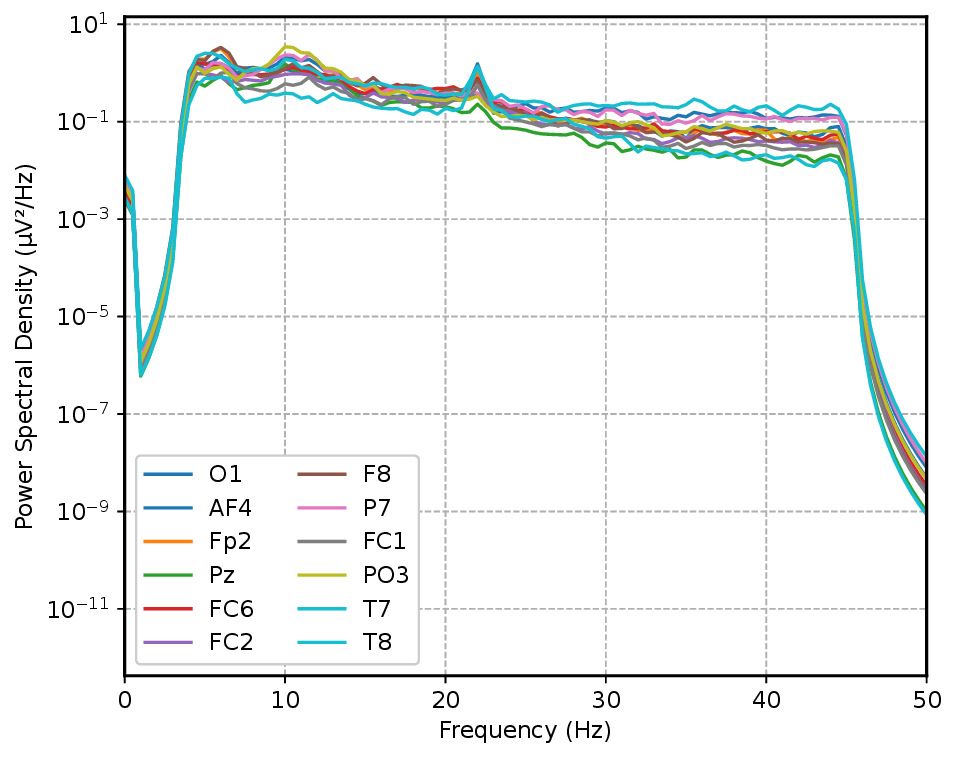}}\hfill
  \subfloat[]{\includegraphics[width=0.32\linewidth]{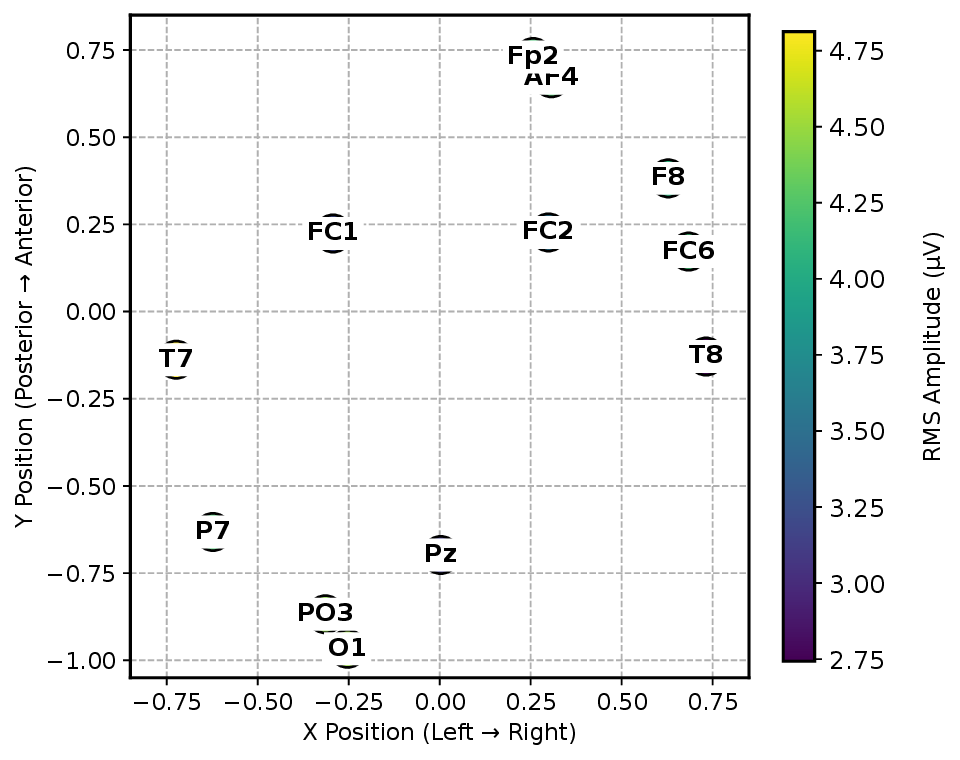}}
  \caption{Multi-domain visualization of the top 12 EEG electrodes:
    (a) temporal signal evolution, (b) frequency-domain PSD profiles,
    and (c) spatial scalp map. The visualizations jointly
    characterize the temporal, spectral, and spatial properties of
    discriminative EEG features.}
  \label{multi-domain}
\end{figure}

To interpret the model's predictions and identify influential EEG
features, we use SHapley Additive exPlanations
(SHAP)~\cite{lundberg2017unified}. Fig.~\ref{SHAP} shows SHAP value
distributions for representative subjects. SHAP values are computed
for positive and negative emotions, and the top 20 channel--band pairs
(across 32 channels and five frequency bands) are selected based on
the mean absolute SHAP values.

The most influential features concentrate in occipital, prefrontal,
frontal-central, temporal, and parietal regions, with gamma and beta
bands dominating across subjects. A consistent set of influential
channels emerges across experimental runs, including O1, AF4, Fp2,
Pz, FC6, FC2, FC1, F8, F4, T7, T8, PO3, and Oz. Notably, the
occipital channel O1 consistently ranks first. Gamma and beta band
features consistently exhibit high SHAP values, underscoring the role
of higher-frequency neural oscillations in emotion discrimination.
These findings align with prior neurophysiological studies on emotion
processing~\cite{imtiaz2025towards,nakajima2021preserving,mouri2023identifying,park2011emotion}.

We further present a multi-domain analysis using the top 12 electrodes
(Fig.~\ref{multi-domain}). The temporal view reveals discriminative
neural patterns over time; the frequency-domain view presents PSD
profiles; and the spatial view maps electrode locations onto a 2D
scalp layout. Together, these visualizations provide a unified
interpretation of important EEG features in terms of their temporal,
spectral, and spatial characteristics.

The results demonstrate that the proposed method achieves strong and
consistent performance across multiple datasets. Even under
significant distribution shifts and without access to source-domain
data, it reliably outperforms existing state-of-the-art approaches.

% -------------------------------------------------------
\section{Conclusion}
\label{conclusion}

In this work, we propose an enhanced source-free unsupervised domain
adaptation framework for cross-dataset EEG-based emotion recognition.
The method addresses domain shifts, absence of source data, and
unreliable pseudo-labels through a unified multi-stage design. We
introduce a non-contrastive predictive coding--based self-supervised
pretraining strategy to learn robust EEG representations by modeling
temporal dependencies without labeled data. During adaptation,
Multi-Loss Adaptive Regularization and Localized Consistency Learning
jointly improve prediction stability and neighborhood consistency
under noisy supervision. A lightweight test-time training mechanism
further refines uncertain samples, enhancing robustness during
inference. Extensive experiments on the DEAP, SEED, and DREAMER
datasets demonstrate that the proposed approach consistently
outperforms state-of-the-art SF-UDA methods in both binary and
multi-class settings. Future work will explore continual and lifelong
adaptation for evolving target distributions, as well as extensions to
other physiological signals such as ECG and EMG, and multimodal EEG
fusion for more robust affective computing systems.

\section*{Acknowledgments}
This work was supported in part by the Natural Sciences and
Engineering Research Council of Canada (NSERC) and in part by the New
Frontiers in Research Fund (NFRF).

% -------------------------------------------------------
% Inline bibliography (no external .bib needed)
% -------------------------------------------------------

\end{document}